\documentclass[a4paper, 12pt]{article}
\usepackage[dvips]{graphicx}

\def\ltap{\ \raise.3ex\hbox{$<$\kern-.75em\lower1ex\hbox{$\sim$}}\ }
\def\gtap{\ \raise.3ex\hbox{$>$\kern-.75em\lower1ex\hbox{$\sim$}}\ }
\def\gl{\ \raise.4ex\hbox{$>$\kern-.75em\lower1ex\hbox{$<$}}\ }

\begin{document}

\title{$\Lambda^{\ast}$-hypernuclei in phenomenological nuclear forces}
\author{A.~Arai, M.~Oka and S.~Yasui  \\
\normalsize Department of Physics, Tokyo Institute of Technology,\\
\normalsize Tokyo 152-8551, Japan \\}
\maketitle

\begin{abstract}
The $\Lambda^{\ast}$-hypernuclei, which are bound states of $\Lambda(1405)$ and nuclei, are discussed as a possible interpretation of the $\bar{K}$-nuclei.
The Bonn and Nijmegen potentials are extended and used as a phenomenological potential between $\Lambda^{\ast}$ and $N$.
The $K$-exchange potential is also considered in the $\Lambda^{\ast}$ and $N$ interaction.
The two-body ($\Lambda^{\ast}N$) and three-body ($\Lambda^{\ast}NN$) systems are solved by a variational method.
It is shown that the spin and isospin of the ground states are assigned as $\Lambda^{\ast}N(S=1, I=1/2)$ and $\Lambda^{\ast}NN(S=3/2, I=0)$, respectively.
The binding energies of the $\Lambda^{\ast}$-hypernuclei are discussed in comparison with experiment.
\end{abstract}

\section{Introduction}

Possibility of kaon-nuclear bound states is of great interest in hadron and
nuclear physics.
A deeply bound kaonic nuclear state with a relatively small decay width
was recently predicted and further studied in literatures.\cite{Akaishi_Yamazaki_02,Dote_etal_04,Dote_etal_04b,Dote_Weise_07} \ 
Possible high density matter caused by the strong kaon attractive force
is also a subject of heated discussion.\cite{Akaishi_Yamazaki_02,Dote_etal_04,Dote_etal_04b,Dote_Weise_07} \ 
In order to find such exotic states, several experimental searches have
been carried out.\cite{Iwasaki_04,FINUDA_05} \ 
FINUDA collaboration has reported an observation of bound $ppK^-$
state at the binding energy, 115 MeV.\cite{FINUDA_05} \ 
Interpretation of the peak found in this experiment is yet under
discussion,
while some advanced calculations, such as a 3-body calculation \`a la
Fadeev equation with $\bar K N - \pi \Sigma$ channel coupling, have
been performed.\cite{Shevchenko_etal_07,Ikeda_Sato_07} \ 
Other approaches include an interpretation as a nine-quark state
studied
in the MIT bag model \cite{Maezawa_etal_05}, and a kaon absorption process between two nucleons \cite{Magas_etal_06}.
While the coming J-PARC facility will certainly answer to the question
whether
such deeply bound kaonic nuclear states exist, we need to study such a system
in more extensive views.

The purpose of this paper is to give a new interpretation to the
``kaon-nucleus''
states.  We consider $\Lambda^*$-hypernuclear states.
$\Lambda^*$ is the lowest negative parity baryon with mass around 1405 MeV
and strangeness $-1$.
This baryon is in many senses unique.
For instance, it is below any other non-strange baryons
with negative parity and is isolated not forming an octet in $SU(3)$.
If it is assumed to be a $p$-wave baryon with spin 1/2,
then it requires a large spin-orbit splitting.
Its uniqueness has attracted a lot of attention and various exotic views
of $\Lambda^*$ have been proposed.

We here do not consider specific composition of $\Lambda^*$, but simply
assume
that it belongs to a flavor singlet representation, and thus isolated.
Our claim is that the so-called kaon-nuclear bound states can be
interpreted as
a bound state of $\Lambda^*$ in a nucleus, or $\Lambda^*$-hypernucleus.
Then the FINUDA observation is regarded as a two-body bound state of
$\Lambda^*$ and $N$, whose binding energy is about 88 MeV.
Larger systems, such as strange tribaryon can be a $\Lambda^* NN$ system
and so on.

In this paper, we construct a model of $\Lambda^* N$ interaction
according to
the one-boson exchange model and consider the two-body ($\Lambda^* N$) and
three-body ($\Lambda^* NN$) systems.
As the $NN$ interaction, we choose the same one-boson exchange model,
the Bonn potential \cite{Bonn} and the Nijmegen soft-core potential models \cite{SC89,ESC04}.
We extend these models to the $\Lambda^* N$ systems and
the $K$-exchange is also included in the same context between $\Lambda^*$
and $N$.
We then solve the two-body
and three-body Schr\"odinger equations using a variational method.

The content of the paper is as follows.  In Section 2, we construct
the potential model for $\Lambda^* N$ system.
The features of the model in the context of possible quantum numbers
of the $\Lambda^*$-nuclear states are also given.
In Section 3, the numerical results for the binding energies of two-body
and three-body bound states are shown.
In Section 4, some discussion on the present results are given.
The conclusion is given in Section 5.

\section{Model}

The phenomenological nuclear force is quite successful in studying nuclear systems.
In order to setup the model for the $\Lambda^{\ast}$-hypernuclei, we discuss the phenomenological $\Lambda^{\ast}N$ interaction. 
The nuclear forces among hyperons are not yet established, although a great amount of studies have been performed experimentally and theoretically so far.
In general, it is considered that the interaction between baryons is described by exchange of mesons supplemented by phenomenological short-range repulsion.
In the present study, based on the one-boson exchange picture, we extend the phenomenological nuclear forces between the $NN$ pair to the $\Lambda^{\ast}N$ pair interaction.
In the Bonn \cite{Bonn} and Nijmegen potentials \cite{SC89,ESC04}, the scalar ($\sigma$, $a_{0}$), pseudoscalar ($\pi$, $\eta$) and vector ($\omega$, $\rho$) bosons are exchanged between the $NN$ pair.
The explicit equations and the parameter sets for the Bonn potential are shown in Appendix.
By considering the $SU(3)$ symmetry, we assume that these potentials are also applied to the $\Lambda^{\ast}N$ pair.
Here the isovector mesons ($a_{0}$, $\pi$, $\rho$) are irrelevant to the $\Lambda^{\ast}N$ interaction, since the $\Lambda^{\ast}$ is isosinglet.
Instead the $K$ (and $\bar{K}$) meson comes into the game in the exchange process $\Lambda^{\ast}N \rightarrow N\Lambda^{\ast}$.
Then, the $\Lambda^{\ast}N$ potential is given by the sum
\begin{eqnarray}
  V_{\Lambda^{\ast}N} = V_{\sigma} + V_{\eta} + V_{\omega} + V_{K}.
\end{eqnarray}
The $\sigma$-exchange potential, $V_{\sigma}$, is the most attractive force, and the $\omega$-exchange potential, $V_{\omega}$, plays an essential role for the spin determination, as we see below.
In the following, we investigate several possibilities of the $\Lambda^{\ast}N$ potential.

In the extended phenomenological nuclear forces in the $\Lambda^{\ast}N$ pair, we have some unknown parameters; the coupling constants $g_{\Lambda^{\ast}NM}$ (for $M$ meson) and the momentum cutoffs, which appear in the form factors.
Among them, the $\Lambda^{\ast}N\bar{K}$ vertex constant, $g_{\Lambda^{\ast}N\bar{K}}$, is determined by the $SU(3)$ relation from the observed decay width of the $\Lambda^{\ast} \rightarrow \Sigma \pi$ channel.
The $\Lambda^{\ast} \Sigma \pi$ coupling $g^{2}_{\Lambda^{\ast} \Sigma \pi}/4\pi = 0.064$ is obtained from the decay width $\Gamma(\Lambda^{\ast} \rightarrow \Sigma \pi)=50.0 \pm 2.0$ MeV.
Assuming that $\Lambda^{\ast}$ is a purely $SU(3)$ singlet state, the $SU(3)$ isoscalar factor,
\begin{eqnarray}
 (\Lambda^{\ast}) = ( N \bar{K} \mbox{ } \Sigma \pi \mbox{ } \Lambda \eta \mbox{ } \Xi K)
  = \frac{1}{\sqrt{8}} ( 2  \mbox{ } 3  \mbox{ } -1  \mbox{ } -2 )^{1/2},
\end{eqnarray}
predicts the coupling constant $g^{2}_{\Lambda^{\ast} N K}/4\pi = g^{2}_{\Lambda^{\ast} \Sigma \pi}/4\pi = 0.064$.
It should be noted that this coupling constant is much smaller than the $N N \sigma$ coupling constant $g^{2}_{N N \sigma}/4\pi=7.78$.
Therefore, the $K$ meson plays only a minor role in the $\Lambda^{\ast}N$ bound state as seen in the numerical calculation later.

Here, the $K$-exchange has to be treated carefully.
Because $\Lambda^*$ lies close to the $\bar K N$ threshold,
the kaon may have four momentum, $q_{\mu}$, near
the on-mass-shell values.  Then the kaon propagator can be enhanced
strongly.  In order to include this effect, we replace the kaon mass
by an effective mass.  Assuming that the baryons are static, we obtain
the finite energy transfer, $q_0 \simeq M_{\Lambda^*} - M_N$, and
the effective mass,
\begin{eqnarray}
&& \tilde m_K \equiv \sqrt{m_k^2- q_0^2}\simeq \sqrt{m_K^2
-(M_{\Lambda^*}-M_N)^2} = 171 \, {\rm MeV}/c^2.
\label{eq:onshell}
\end{eqnarray}
This replacement indeed enhances the $\Lambda^* N$ potential more
strongly than the case of
$\Lambda N$ or $\Sigma N$ potential.
This is because that the $\Lambda^* N \bar K$ coupling is a scalar
coupling and the
effect of $\tilde m_K$ comes only in the range of the potential.
In contrast, a $p$-wave coupling gives rise to an extra $\tilde m_K^2$
factor,
which significantly suppresses its effect on the $\Lambda N$ or $\Sigma
N$ potential.
As a result, the $K$-exchange potential for $\Lambda^{\ast}N$ pair is given by the coupling constant $g^{2}_{\Lambda^{\ast} N K}/4\pi = 0.064$ and the effective mass $\tilde{m_{K}}$ in Eq.~(\ref{eq:pspot}) in Appendix.

Yet the other parameters in the $\Lambda^{\ast}N$ interaction are not fixed due to lack of experimental information.
In the present study, we treat the $g_{\Lambda^{\ast} \Lambda^{\ast} \sigma}$ as a free parameter, since the results happen to be most sensitive to the $\sigma$-exchange.
We further assume that the other parameters are the same as the $NN$ interaction for simplicity.

Here it is important for the later discussions to prospect the properties of the $\Lambda^{\ast}N$ interaction.
Let us see the spin-dependence of the $\Lambda^{\ast}N$ interaction.
The spin-spin interaction is induced by the $\eta$, $K$ and $\omega$ mesons.
We consider the $\omega$-exchange force, since the $\omega$ meson coupling is much stronger than the $\eta$ and $K$ meson couplings.
The static $\omega$-exchange potential is written in the momentum space as
\begin{eqnarray}
 V_{\omega}(\vec{k}^{2}) &=& -C \frac{ (\vec{\sigma}_{1} \cdot \vec{\sigma}_{2}) \vec{k}^{2} }{\vec{k}^{2}+m^{2}_{\omega}}
 \nonumber \\
             &=& -C \left( 1- \frac{m^{2}_{\omega}}{\vec{k}^{2}+m^{2}_{\omega}} \right) (\vec{\sigma}_{1} \cdot \vec{\sigma}_{2}),
 \label{eq:spinspin}
\end{eqnarray}
with a positive constant $C$, and the $\omega$ meson mass $m_{\omega}$.
In the real space, the first term is a delta function type, while the second term is the standard Yukawa potential.
With a form factor introduced, the delta type potential becomes a finite range potential.
For the Yukawa type potential, the spin singlet is more attractive than the spin triplet.
On the other hand, for the delta type potential, the spin triplet is more attractive than the spin singlet.
Usually the delta type potential at short distance plays a minor role due to strong repulsive core of the $NN$ force, hence the Yukawa potential is much more important.
However, this is not the case for the $\Lambda^{\ast}N$ potential.
There, at short distances, the delta function is overwhelming to the Yukawa type potential.
Consequently, in the $\Lambda^{\ast}N$ potential, the spin triplet potential is more attractive at short distance than the spin singlet one.

Now we look at the possible quantum states of the $\Lambda^{\ast}$-hypernuclei, which are obtained in the spin and isospin combinations.
It is assumed that all the particles occupy the lowest-energy $s$-orbit.
Since $\Lambda^{\ast}$ has spin $S=1/2$ and isospin $I=0$, the $\Lambda^{\ast}N$ system has two states of spin and isospin; $(S=0, I=1/2)$ and $(S=1, I=1/2)$.
On the other hand, the $\Lambda^{\ast}NN$ system takes the isosinglet and isotriplet states.
The isosinglet state has the spin $S=1/2$ and $S=3/2$ states, where the $NN$ pair is isosinglet and spin triplet.
The isotriplet state has spin $S=1/2$, where the $NN$ pair is isotriplet and spin singlet.
Therefore the possible quantum numbers of the $s$-wave $\Lambda^{\ast}NN$ system are $(S=1/2, I=0)$, $(S=3/2, I=0)$ and $(S=1/2, I=1)$.

Here, we prospect the quantum numbers of the ground states of the $\Lambda^{\ast}$-hypernuclei by using the spin dependence of the $\omega$-exchange at short distance.
It is expected that the ground state of the two body system ($\Lambda^{\ast}N$) is spin triplet.
In order to understand the ground state of the three body system ($\Lambda^{\ast}NN$), the three possible states, $(S=1/2, I=0)$, $(S=3/2, I=0)$ and $(S=1/2, I=1)$, are expanded in terms of the $\Lambda^{\ast}N$ pair; 
\begin{eqnarray}
 | \Lambda^{\ast} (NN)_{S=1}; S=1/2, I=0 \rangle
  &=&  \frac{1}{2} | (\Lambda^{\ast}N)_{S=1} N \rangle + \frac{\sqrt{3}}{2} | (\Lambda^{\ast}N)_{S=0} N \rangle,
  \\
 | \Lambda^{\ast} (NN)_{S=1}; S=3/2, I=0 \rangle
  &=&  | (\Lambda^{\ast}N)_{S=1} N \rangle,
  \label{eq:expand1}
\end{eqnarray}
for the isosinglet states, and
\begin{eqnarray}
 | \Lambda^{\ast} (NN)_{S=0}; S=1/2, I=1 \rangle
  &=&  \frac{\sqrt{3}}{2} | (\Lambda^{\ast}N)_{S=1} N \rangle - \frac{1}{2} | (\Lambda^{\ast}N)_{S=0} N \rangle,
  \label{eq:expand2}
\end{eqnarray}
 for the isotriplet state.
Among these states, the spin triplet interaction is contained most strongly in the $(S=3/2, I=0)$ state due to the largest value of coefficients of the  $(\Lambda^{\ast}N)_{S=1}$ component.
Therefore, it is expected that the $(S=3/2, I=0)$ is the ground state for the $\Lambda^{\ast}NN$ state.
In the next section, it will be shown that the above analysis is consistent with the numerical results.

The above conclusion of the spin of the two-body and three-body states is in strong contrast with the kaon bound state approach.
There, the spin dependence is induced by the isospin dependence of $\bar{K}N$ interaction.
For instance, in the $\bar{K}NN$ system, the $(\bar{K}N)_{I=0}$ interaction is the driving force for the bound state.
It is easy to see that the $\bar{K}(NN)_{I=1, S=0}$ state contains more $(\bar{K}N)_{I=0}$ component than $\bar{K}(NN)_{I=0, S=1}$ state.
Thus, the ground state of $\bar{K}NN$ is expected to have $S=0$.

\section{Numerical Results}

In this section, we discuss the numerical results for the two-body ($\Lambda^{\ast}N$) and three-body ($\Lambda^{\ast}NN$) systems by solving the Schr\"odinger equation with the Bonn \cite{Bonn} and Nijmegen (SC89 \cite{SC89} and ESC04 \cite{ESC04}) potentials for the $\Lambda^{\ast}N$ and $NN$ interactions.
For simplicity, derivative terms are not considered.
We mainly show the results for the Bonn potential.
The results do not differ qualitatively for the Nijmegen potentials with SC89 and ESC04.

\subsection{Two-body system}

\begin{figure}[tbp]
 \begin{center}
 \includegraphics[width=8cm]{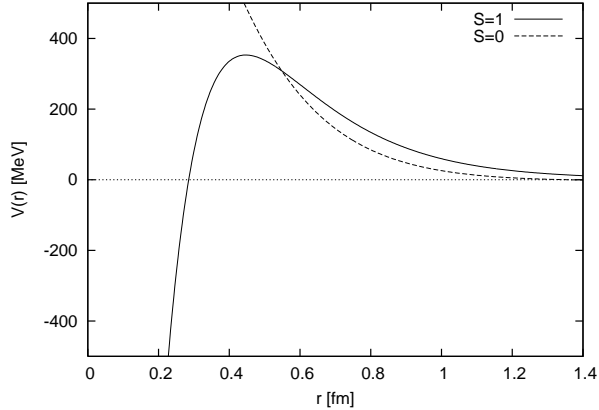} \\
 \end{center}
  \caption{\small \baselineskip=0.5cm The Bonn potentials applied for $\Lambda^{\ast}N$ pair in the $S=0$ (dashed line) and $S=1$ (solid line) channels for the coupling constants $g_{\Lambda^{\ast} \Lambda^{\ast} \sigma}/g_{NN \sigma}$=0.39.}
  \label{fig:BonnYNs0s1-39}
\end{figure}

The two-body system ($\Lambda^{\ast}N$) has the spin singlet $(S=0, I=1/2)$ and triplet $(S=1, I=1/2)$ states.
In Fig.~\ref{fig:BonnYNs0s1-39}, the Bonn potentials are plotted as functions of the relative distance $r$ between $\Lambda^{\ast}$ and $N$ in the $S=0$ and $S=1$ channels for the coupling constant $g_{\Lambda^{\ast} \Lambda^{\ast} \sigma}/g_{NN \sigma}=0.39$.
The $S=1$ potential is strongly attractive at short distance ($r \ltap 0.4$ fm), while the $S=0$ one is repulsive.
This difference is caused by the delta type interaction in the $\omega$ potential (\ref{eq:spinspin}), which is attractive for $S=1$ and repulsive for $S=0$.
It should be noted that the delta function is smeared to a finite range potential due to the form factor in the phenomenological nuclear forces.

\begin{figure}[tbp]
 \begin{center}
 \includegraphics[width=8cm]{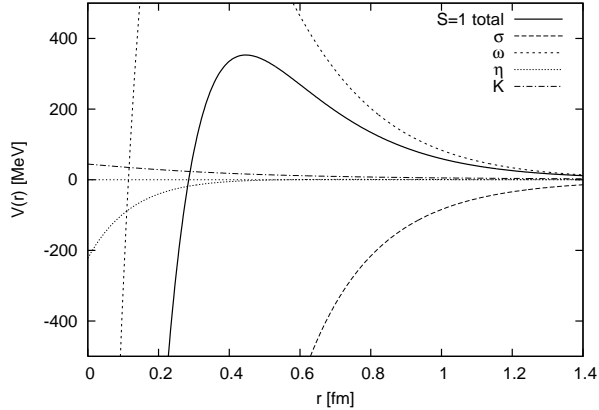} \\
 \end{center}
   \caption{\small \baselineskip=0.5cm The components from the $\sigma$, $\omega$, $\eta$ and $K$ mesons in the Bonn potential between $\Lambda^{\ast}N$ pair for $S=1$ for $g_{\Lambda^{\ast} \Lambda^{\ast} \sigma}/g_{NN \sigma}$=0.39. The Bonn potential is indicated by the solid line.}
   \label{fig:BonnYNs1-39}
\end{figure}

In order to see the contributions from individual mesons in the $S=1$ $\Lambda^{\ast}N$ potential, the components from the $\sigma$, $\omega$ and $\eta$-exchange potentials and the additional $K$-exchange potentials are plotted in Fig.~\ref{fig:BonnYNs1-39}.
The $\sigma$-exchange potential is attractive at medium range ($r \gtap 0.4$ fm), while the $\omega$ potential is repulsive.
However, the $\omega$ potential is strongly attractive at shorter distances ($r \ltap 0.4$ fm) due to the delta function in (\ref{eq:spinspin}).
Hence, the sum of them employs a deeply attractive potential at short distance, which is one of the characteristic properties in the $\Lambda^{\ast}N$ potential.
The contributions from the $\eta$ and $K$ mesons are relatively weak.

\begin{figure}[tbp]
 \begin{center}
 \includegraphics[width=8cm]{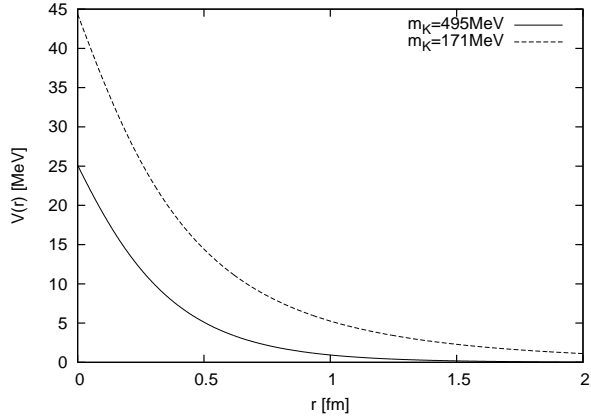} \\
 \end{center}
   \caption{\small \baselineskip=0.5cm The $K$-exchange component in the $\Lambda^{\ast}N$ potential. The solid line is for the bare mass $m_{K}=495$ MeV and the dashed line is for the reduced mass $\tilde{m}_{K}=171$ MeV. See the text.}
   \label{fig:Bonnkaon}
\end{figure}

Here we investigate the on-shell effect of the $K$-exchange.
As we have already discussed, the $K$ meson propagating between the $\Lambda^{\ast}$ and $N$ may have large energy and thus can be close to the on-mass-shell kinematics.
Such $K$-exchange might be largely enhanced.
We treat this effect as an effective mass $\tilde{m}_{K}$ defined in Eq.~(\ref{eq:onshell}).
In Fig.~\ref{fig:Bonnkaon}, the $K$-exchange potential is plotted for the bare mass $m_{K}=495$ MeV (solid line) and for the reduced mass $\tilde{m}_{K}=171$ MeV (dashed line).
As the effective $K$ meson mass gets smaller, both the potential strength and the range increase.
However, this enhancement has little significance to the binding energy of the $\Lambda^{\ast}N$, since the $K$-exchange potential is still weaker than the $\sigma$ and $\omega$-exchange potentials.

\begin{figure}[tbp]
 \begin{center}
 \includegraphics[width=8cm]{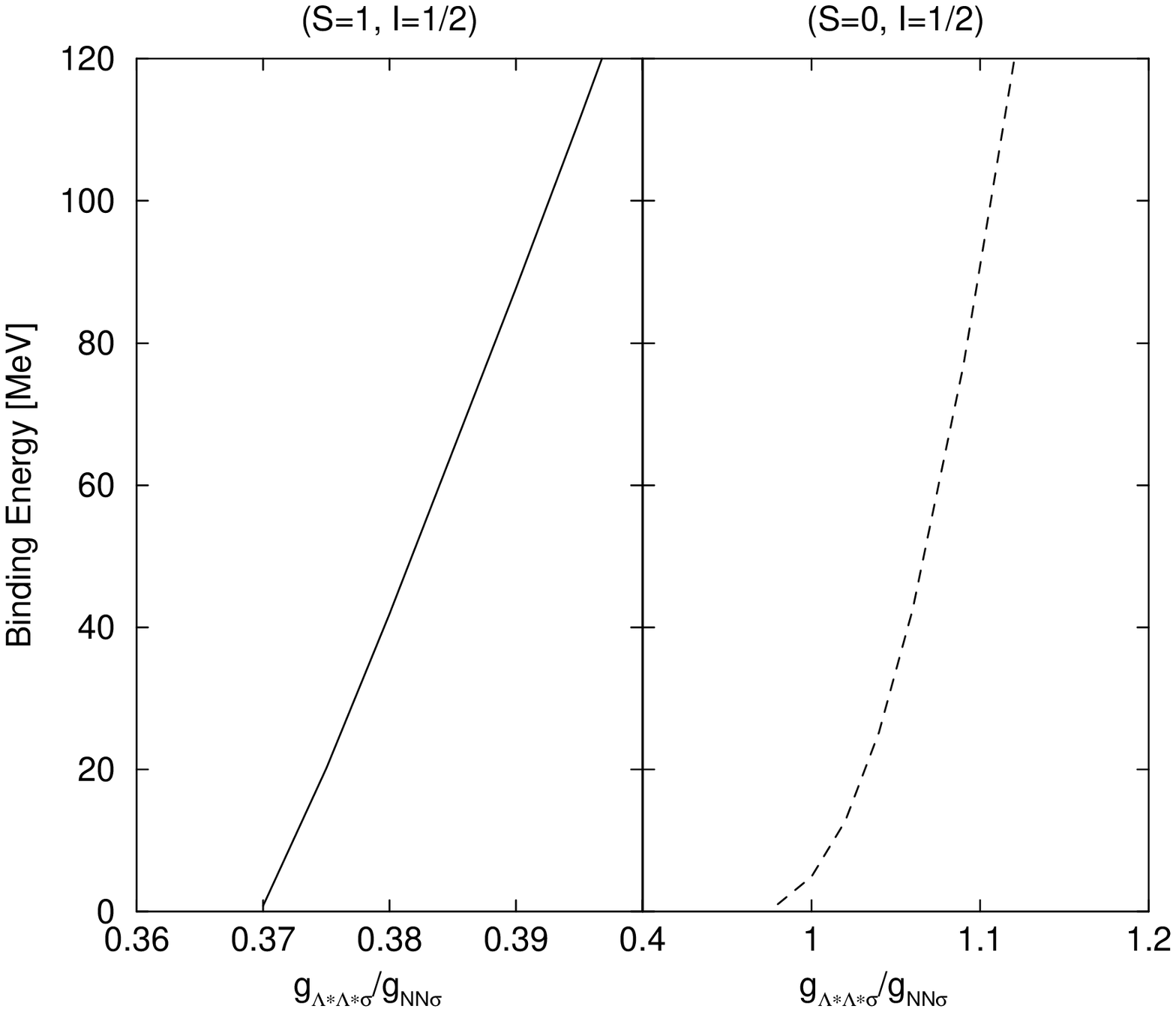} \\
 \end{center}
   \caption{\small \baselineskip=0.5cm The binding energy of the  $\Lambda^{\ast}N$ as functions of the coupling constant $g_{\Lambda^{\ast}\Lambda^{\ast}\sigma}/g_{NN\sigma}$ for the Bonn potential. The solid line is for $S=1$ and the dashed line for $S=0$.}
   \label{fig:Bonnenen}
\end{figure}

Solving the Schr\"odinger equation for the $\Lambda^{\ast}N$ with the Bonn potential, we obtain the binding energy and the wave function of the $\Lambda^{\ast}N$ bound states.
As a set of trial wave functions, we use a linear combination of
Gaussian functions;
\begin{eqnarray}
\psi(r) = \sum_{i=1}^{n} a_{i} e^{-\alpha_{i} r^{2}},
\end{eqnarray}
where $a_{i}$ are the normalization constants, $\alpha_{i}$ the width parameters, and $r$ the relative distance between $\Lambda^{\ast}$ and $N$.
As a result, the binding energy strongly depends on the coupling constant $g_{\Lambda^{\ast}\Lambda^{\ast}\sigma}/g_{NN\sigma}$ as shown in Fig.~\ref{fig:Bonnenen}.
The $\sigma$ potential plays an essential role for the formation of the bound state.
Indeed, without the $\Lambda^{\ast} \Lambda^{\ast} \sigma$ coupling, the bound state does not exist.
For $S=1$, the $\Lambda^{\ast}N$ is bound for the coupling constant $g_{\Lambda^{\ast}\Lambda^{\ast}\sigma}/g_{NN\sigma} \gtap 0.37$.

\begin{figure}[tbp]
 \begin{center}
 \includegraphics[width=8cm]{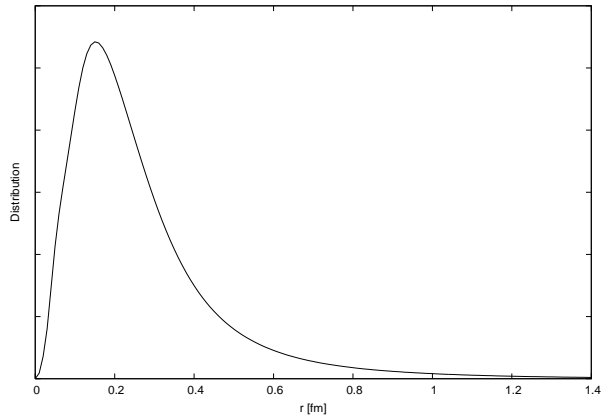} \\
 \end{center}
   \caption{\small \baselineskip=0.5cm The probability density of the bound $\Lambda^{\ast}N$ system with $S=1$ and $g_{\Lambda^{\ast}\Lambda^{\ast}\sigma}/g_{NN\sigma}=0.39$. The binding energy is 88 MeV. The horizontal axis is the relative distance between $\Lambda^{\ast}$ and $N$.}
   \label{fig:wavYNs1Bonn}
\end{figure}

At $g_{\Lambda^{\ast}\Lambda^{\ast}\sigma}/g_{NN\sigma} = 0.39$, the binding energy reaches 88 MeV.
The observed binding energy 115 MeV of $ppK^{-}$ state reported by the FINUDA collaboration \cite{FINUDA_05} is interpreted as 88 MeV for the binding energy of the $\Lambda^{\ast}$N state.
For $S=0$, a stronger coupling $g_{\Lambda^{\ast}\Lambda^{\ast}\sigma}/g_{NN\sigma} \gtap 0.98$ is required for the bound state.
The binding energy 88 MeV is obtained at $g_{\Lambda^{\ast}\Lambda^{\ast}\sigma}/g_{NN\sigma} = 1.01$.
In the current experimental status, the quantum number of the bound state is not yet known.
Our result suggests an $S=1$ bound state rather than $S=0$.
The probability density of the $\Lambda^{\ast}N$ with $S=1$ is plotted in Fig.~\ref{fig:wavYNs1Bonn}.
The wave function is very compact as compared with the nucleon size, suggesting that the obtained bound state is produced mainly by the short-range $\omega$ attraction assisted by the medium-range $\sigma$-exchange that cancels the medium range $\omega$ repulsion.

So far, we have discussed the results of the Bonn potential.
In order to check the model dependence, we also apply different types of the
nuclear force.
In particular, as the short-range part of the interaction is important,
the mechanisms of short-range $NN$ repulsion is in question.
The Bonn potential acquires the short-range repulsion mainly from the
$\omega$-exchange potential, while the other models, such as the Nijmegen potential,
introduce a new component for the repulsion.

The pomeron exchange is represented in the Nijmegen potential by a gaussian potential of short range.
In the case of $NN$ interaction, it is strong enough to expel the wave functions away from the center.
When we apply the same repulsion to $\Lambda^{\ast}N$ system,
we find that the short-range attraction from $\omega$-exchange may still make a bound state,
 if the $\sigma$-exchange attraction at the medium range is strong enough.
However, compared with the Bonn potential, the extra pomeronic repulsion for $\Lambda^{\ast}N$ makes the system less bound.
That means that the required $g_{\Lambda^{\ast}\Lambda^{\ast}\sigma}$ coupling constant becomes larger.

We employ the Nijmegen potential of the versions SC89 and ESC04.
It is found that SC89 requires the minimal coupling constants $g_{\Lambda^{\ast}\Lambda^{\ast}\sigma}/g_{NN\sigma}=0.77$ and $g_{\Lambda^{\ast}\Lambda^{\ast}\sigma}/g_{NN\sigma}=0.98$ to form a $\Lambda^{\ast}N$ bound state for $S=1$ and $S=0$, respectively. 
The binding energy (88 MeV) reported by the FINUDA group is obtained by setting $g_{\Lambda^{\ast}\Lambda^{\ast}\sigma}/g_{NN\sigma}=0.832$ for $S=1$ and $g_{\Lambda^{\ast}\Lambda^{\ast}\sigma}/g_{NN\sigma}=1.128$ for $S=0$.
The probability density is pushed out from the center as compared with the Bonn potential.
This is because the range of the $\Lambda^{\ast}N$ potential is longer than that from the Bonn potential.

The result of ESC04 is qualitatively the same as that of SC89.
Only difference is that absolute values of the $\sigma$, $\omega$ and pomeron exchange potentials in ESC04 are smaller than those in SC89.
Then, the minimum coupling constants of $\Lambda^{\ast} \Lambda^{\ast} \sigma$ become larger, and the experimental binding energy is reproduced at $g_{\Lambda^{\ast}\Lambda^{\ast}\sigma}/g_{NN\sigma}=1.118$ for $S=1$ and  $g_{\Lambda^{\ast}\Lambda^{\ast}\sigma}/g_{NN\sigma}=1.398$ for $S=0$.

The results above discussed are summarized in Table \ref{tbl:table1}.
The ratios of coupling constants $g_{\Lambda^{\ast}\Lambda^{\ast}\sigma}/g_{NN\sigma}$ that correspond to the experimental observation in the FINUDA collaboration are listed.
The smaller value indicates that the system can form a bound state more easily.
We find that in all cases that the coupling constant in $S=1$ bound state is smaller than that in $S=0$ one in the Bonn and Nijmegen potentials.
Therefore, we expect that the lowest-energy bound state is the $S=1$ state.

\begin{table}[tdp]
\caption{The ratios of the coupling constants $g_{\Lambda^{\ast}\Lambda^{\ast}\sigma}/g_{NN\sigma}$ required to obtain the binding energy, 88 MeV, of  the $\Lambda^{\ast}N$ two-body system.}
\begin{center}
\begin{tabular}{c|c|c|c|c}
\hline
\hline
$\Lambda^{\ast}N$ & Bonn & Nijmegen (SC89) & Nijmegen (ESC04) & B.E. [MeV] \\
\hline
$S=1$    & 0.39 & 0.832  &1.118 & 88 \\
$S=0$   & 1.1 & 1.128  &  1.398 & 88 \\
\hline
\end{tabular}
\end{center}
\label{tbl:table1}
\end{table}%

\subsection{Three-body system}

\begin{figure}[tbp]
 \begin{center}
 \includegraphics[width=8cm]{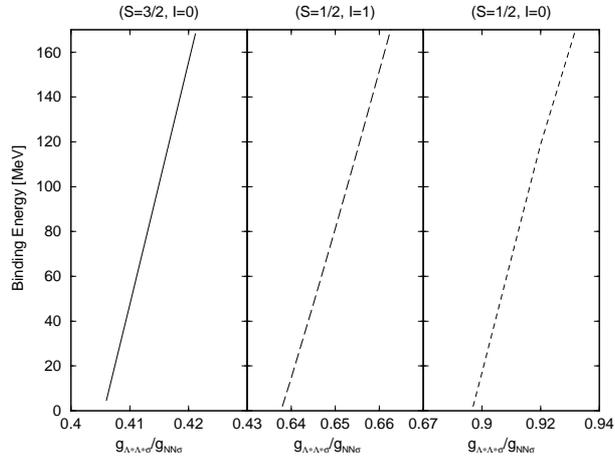} \\
 \end{center}
   \caption{\small \baselineskip=0.5cm The binding energy of the  $\Lambda^{\ast}NN$ as functions of the coupling constant $g_{\Lambda^{\ast}\Lambda^{\ast}\sigma}/g_{NN\sigma}$ for the Bonn potential. The solid, long-dashed and short-dashed lines indicate the $(S=3/2, I=0)$, $(S=1/2, I=1)$ and $(S=1/2, I=0)$ states, respectively.}
   \label{fig:Bonn3ene}
\end{figure}

Now we discuss the three-body system ($\Lambda^{\ast}NN$).
The binding energies and wave functions are obtained variationally by solving the Schr\"odinger equation by using the phenomenological  $NN$ and $\Lambda^{\ast}N$ potentials.
The trial wave function is given by
\begin{eqnarray}
\psi(r, r') = \sum_{i,j} a_{ij} e^{-\alpha_{i} r^{2}} e^{-\beta_{i} r'^{2}}
\end{eqnarray}
with $a_{ij}$ are the normalization constants, $\alpha_{i}$ and $\beta_{i}$ the width parameters, and $r$ and $r'$ the Jacobi coordinates for 
$\Lambda^{\ast}NN$.
Then we obtain the binding energies as functions of the coupling constant $g_{\Lambda^{\ast}\Lambda^{\ast}\sigma}/g_{NN\sigma}$ for the $(S=3/2, I=0)$, $(S=1/2, I=1)$ and $(S=1/2, I=0)$ states indicated by solid, long-dashed, short-dashed lines, respectively, in Fig.~\ref{fig:Bonn3ene}.
We find that a bound $(S=3/2, I=0)$ state can be formed only for $g_{\Lambda^{\ast}\Lambda^{\ast}\sigma}/g_{NN\sigma} \gtap 0.406$, while the $(S=1/2, I=1)$ and $(S=1/2, I=0)$ bound states require stronger coupling constant.
Therefore our analysis shows the $(S=3/2, I=0)$ state as the ground state of the $\Lambda^{\ast}NN$ system.
Note that for $\Lambda^{\ast}N$ pair in the $S=1$ channel is more attractive than the $S=0$ channel.
From the $\Lambda^{\ast}NN$ states written as combinations of $(\Lambda^{\ast}N)_{S=1}N$ and $(\Lambda^{\ast}N)_{S=0}N$ as Eq.~(\ref{eq:expand1}) and (\ref{eq:expand2}), one sees that the $(S=3/2, I=0)$ state contains more the $(\Lambda^{\ast}N)_{S=1}$ component.

Through the discussions given above, it is clear that the short-distance behavior of the nuclear potential plays an important role.
We recall that at short distance the smeared delta interaction in the $\omega$-exchange in $\Lambda^{\ast}N$ interaction overwhelms the Yukawa type potential, and it induces a strong attractive potential for $S=1$.
This is contrary to the long range behavior of the $\omega$-exchange potential; repulsive for $S=1$ and attractive for $S=0$.
The strong attractive potential at short distance is also seen in the three-body system.
It is directly observed from the probability densities of $\Lambda^{\ast}N$ and $\Lambda^{\ast}NN$, which concentrate at short distances.

\begin{table}[tdp]
\caption{The ratios of the coupling constants $g_{\Lambda^{\ast}\Lambda^{\ast}\sigma}/g_{NN\sigma}$ required to obtain the binding energy, 167 MeV, of  the $\Lambda^{\ast}NN$ three-body system.}
\begin{center}
\begin{tabular}{c|c|c|c|c}
\hline
\hline
$\Lambda^{\ast}NN$ & Bonn & Nijmegen (SC89) & Nijmegen (ESC04) & B.E. [MeV] \\
\hline
$(S=3/2, I=0)$ & 0.424 & 0.898  & 1.217 & 167 \\
$(S=1/2, I=1)$ & 0.64   & 0.987  & 1.325 & 167 \\
\hline
\end{tabular}
\end{center}
\label{tbl:table2}
\end{table}%

\section{Discussion}
We here compare our results with some other studies for the same system.
In the original Akaishi-Yamazaki
picture~\cite{Akaishi_Yamazaki_02,Dote_etal_04,Dote_etal_04b},
the binding energy of $ppnK^{-}$ is shown to be larger than $ppK^{-}$.
Furthermore, it is indicated
that $\bar{K}$-nuclei can be stable for larger baryon numbers.
However, this is not the case for the $\Lambda^{\ast}$-hypernuclei.
Let us compare the minimum coupling constants for the two-body ($\Lambda^{\ast}N$) and three-body ($\Lambda^{\ast}NN$) bound systems.
For example, for the Bonn potential, the coupling $g_{\Lambda^{\ast} \Lambda^{\ast} \sigma}/g_{NN\sigma}=0.37$ gives a bound state of $\Lambda^{\ast}N$ with $(S=1, I=1/2)$, while $g_{\Lambda^{\ast} \Lambda^{\ast} \sigma}/g_{NN\sigma}=0.406$ is required for $\Lambda^{\ast}NN$ with $(S=3/2, I=0)$.
Therefore, the present framework leads to the conclusion that the $\Lambda^{\ast}N$ two-body state is bound more easily than the $\Lambda^{\ast}NN$ three-body state.
This is qualitatively the same for the Nijmegen potentials with SC89 and ESC04.
Furthermore, it may be expected that the the $\Lambda^{\ast}$-hypernuclei becomes unbound as the baryon number increases.
This picture is in opposite direction as compared with the $\bar{K}$-nucleus bound states.\cite{Akaishi_Yamazaki_02,Dote_etal_04,Dote_etal_04b}

A recent proposal~\cite{Yamazaki:2007cs} of interpreting
the binding of $\bar K$ in terms of ``migration'' of $\bar K$ is in fact
quite similar to the $K$-exchange part of our picture.
The difference between their analysis and ours is mainly in the coupling
strengths.
We determine the coupling constant so as to reproduce the relatively
narrow width of $\Lambda^*$, while
Ref.~\cite{Yamazaki:2007cs} employs the coupling to
reproduce the binding energy of
$\Lambda^*$ as a $\bar K N$ bound state.  A further analysis on this
difference will be needed.

A comment is in order for the dependence of the result on the coupling constant in the phenomenological nuclear potentials for $\Lambda^{\ast} N$.
The coupling constant $g_{\Lambda^{\ast} \Lambda^{\ast} \sigma}$ is taken as a free parameter, and the couplings for the other mesons are fixed to the same value as the $NN$ interaction.
The $\sigma$ potential is the most attractive force among them, and the $\omega$ potential is the second strongest force.
Therefore, as a further step, it is interesting to alter the coupling strength of $\Lambda^{\ast}\Lambda^{\ast}\omega$.
However, concerning the spin of the bound state, the result will not be changed as long as the $\omega$-exchange potential is attractive at short-range.
This is because that the spin of the $\Lambda^{\ast}N$ pair is almost determined by the $\omega$-exchange potential, not by the other mesons.

Lastly, we note that in the present framework only the meson exchange
potential
has been considered as interaction between $\Lambda^{\ast}$ and $N$.
However, the resulting bound states are compact where the substructure
of the baryons
should be important.
In the present analysis, we have introduced the form factors
representing the baryon structure.
It may be important to consider the quark structure of baryons
explicitly so that
the short range baryon-baryon
interactions are correctly taken into account.\cite{Oka:1980ax}
This subject is left for a future work.

\section{Conclusion}

The possibility of the $\Lambda^{\ast}$-hypernuclei is discussed by considering the $\Lambda^{\ast}$ as a compound state in the kaonic-nuclei.
The two-body ($\Lambda^{\ast}N$) and three-body ($\Lambda^{\ast}NN$) systems are investigated by using phenomenological nuclear forces.
Based on the one-boson exchange picture, the Bonn and Nijmegen (SC89 and ESC04) potentials are extended to the $\Lambda^{\ast}N$ interaction.
The $K$-exchange is also included in the extended nuclear forces.
The $g_{\Lambda^{\ast}N\bar{K}}$ coupling is determined from the experimental value of the decay width of the $\Lambda^{\ast}$.
Only $\Lambda^{\ast} \Lambda^{\ast} \sigma$ coupling constant is left as a free parameter with the other parameters fixed.

As a result, for the two-body system ($\Lambda^{\ast}N$), an appropriate $\Lambda^{\ast} \Lambda^{\ast} \sigma$ coupling constant reproduces the binding energy of the $ppK^{-}$ which is comparable to the value reported by the FINUDA collaboration.
The most stable state in the $\Lambda^{\ast}N$ bound state is the $S=1$ state.
For the three-body system ($\Lambda^{\ast}NN$),
the $(S=3/2, I=0)$ state is the most stable state.
These results are understood by the fact that the $\omega$-exchange potential is strongly attractive at short distances for $S=1$ channel rather than in $S=0$.
There the $K$-exchange plays only a minor role.
Compared with the $\Lambda^{\ast}N$ and $\Lambda^{\ast}NN$, the minimum coupling constant $g_{\Lambda^{\ast} \Lambda^{\ast} \sigma}$ in the former is smaller than that of the latter.
Therefore, the $\Lambda^{\ast}N$ state is more easily produced in comparison with the $\Lambda^{\ast}NN$ state.
Our conclusion is qualitatively the same for the Bonn and Nijmegen (SC89 and ESC04) potentials.

In the present study, $\Lambda^{\ast}$ is considered as a stable particle.
In reality, however, multi-channel decays are open for $\Lambda^{\ast}$.
The decay to $\Sigma \pi$ is the main source of $\Lambda^{\ast}$ free decay, while the in-medium decays $\Lambda^{\ast} N \rightarrow \Lambda N$, $\Sigma N$ are new and interesting.
In our discussion, the bound $\Lambda^{\ast}N$ and $\Lambda^{\ast}NN$ states are very compact objects.
Therefore, the conversion width to $\Lambda N$ or $\Sigma N$ may be modified.
These subjects are closely related to experimental researches in DA$\Phi$NE and J-PARC.
Further studied are left for future works.

\section*{Acknowledgement}
This work is supported by a Grant-in-Aid for Scientific Research for Priority Areas, MEXT (Ministry of Education, Culture, Sports, Science and Technology) with No. 17070002.

\appendix
\section{}
The Bonn potential used in the present paper is explicitly shown in a coordinate formalism \cite{Bonn}.
In the followings, $m$ is the mass of the meson, $g_{ij}$ and $f_{ij}$ are coupling constants in the reaction process, $1+2 \rightarrow 3+4$ with the baryon mass $M_1$ and $M_2$.
The parameter set is summarized in Table \ref{tb:Bonn_parameter}.
The pseudoscalar type potential is given by
\begin{eqnarray}
V_{\rm ps} (m, r) = \frac{g_{13}g_{24}}{4\pi} \frac{m^2}{4{M_1} {M_2}} m \left[\frac{1}{3} \vec{\sigma _1} \cdot \vec{\sigma _2} \phi (mr) + S_{12} \chi (mr) \right].
\label{eq:pspot}
\end{eqnarray}
The scalar type potential is given by
\begin{eqnarray}
 V_{\rm s}(m, r) = -\frac{g_{13}g_{24}}{4\pi} \left(1 -  \frac{m^2}{8M_1M_2}\right) \phi (mr).
\label{eq:spot}
\end{eqnarray}
The vector type potential is given by
\begin{eqnarray}
 V_{\rm v}(m, r) =&& \frac{m}{4\pi} \biggl[  \biggl\{ g_{13}g_{24} \left( 1 +  \frac{m^2}{8M_1M_2} \right)  + g_{13}f_{24} \frac{m^2}{4MM_1} + f_{13}g_{24} \frac{m^2}{4MM_2}
 + f_{13}f_{24}\frac{m^4}{16M^2M_1 M_2} \biggl\} \phi(mr)
   \nonumber \\
 &&+\frac{m^2}{4M_1M_2} \biggl\{ g_{13}g_{24} + g_{13}f_{24} \frac{M_2}{M} + f_{13}g_{24} \frac{M_1}{M} + f_{13}f_{24} \frac{M_1M_2}{M^2} 
 \biggl( 1 + \frac{m^2}{8M_1M_2} \biggl) \biggl\} \nonumber \\
&& \left( \frac{2}{3} ( \vec{\sigma _1} \cdot \vec{\sigma _2} ) \phi(mr) - S_{12} \chi(mr) \right) \biggl] .
\nonumber \\
\label{eq:vpot}
\end{eqnarray}
In the above equations, we define
\begin{eqnarray}
 \phi (x) &=& \frac{e^{-x}}{x}, \\ \chi(x) &=& \left( \frac{m}{M} \right)^{2} \left( 1+\frac{3}{x}+\frac{3}{x^{2}} \right) \phi(x), \\ S_{12} &=& 3 \frac{(\vec{\sigma_1} \cdot \vec{r})(\vec{\sigma_2} \cdot \vec{r})}{r^2} - \vec{\sigma}_{1} \cdot \vec{\sigma}_2,
\end{eqnarray}
with the propagating meson mass $m$.
When the exchanged meson has isospin, $\tau_{1} \cdot \tau_{2}$ is multiplied.
The form factor is introduced for each meson by replacing as
\begin{eqnarray}
V_\alpha (m, r) \rightarrow V_\alpha (m , r) - \frac{{\Lambda_{2}}^2 - {m}^2}{{\Lambda_{2}}^2 -{\Lambda_{1}}^2}
  V_\alpha (\Lambda_{1} , r) +  \frac{{\Lambda_{1}}^2 - {m}^2}{{\Lambda_{2}}^2 -{\Lambda_{1}}^2}
  V_\alpha (\Lambda_{2} , r),
\end{eqnarray}
where
\begin{eqnarray}
   \Lambda_{1} = \Lambda + \epsilon , \, \Lambda_{2} = \Lambda - \epsilon,
\end{eqnarray}
with $ \epsilon/ \Lambda \ll 1$, such as $\epsilon \approx 10$ MeV.

\begin{table}[tdp]
\begin{center}
 \caption{The parameter set of the Bonn potential. (The superscript $^a$ and $^b$ denote the $NN$ pairs with isospin $T=1$ and $T=0$, respectively. )}
 \begin{tabular}{lrrrr} \hline
  Meson & Mass (MeV) &$g^2/4 \pi$ & $f^2/4 \pi$ & $\Lambda$ (GeV) \\ \hline
  $\pi$ & 138.03 & \ & 14.9 & 1.3 \\
  $\eta$ & 548.8 & 3 & \ & 1.5 \\
  $\rho$ & 769 & 0.95 & 5.8 & 1.3 \\
  $\omega$ & 782.6 & 20 & \ & 1.5 \\
  $a_0$ & 983 & 2.6713 & \ & 2.0 \\
  $\sigma$ & 550$^a$ & 7.7823$^a$  &\ & 2.0 \\
        \  & 715$^b$ & 16.2061$^b$ &\ & 2.0 \\ \hline
 \end{tabular}
 \label{tb:Bonn_parameter}
\end{center}
\end{table}


\end{document}